\documentclass[12pt,epsf]{article}
\usepackage{graphicx}
\usepackage{epsfig}
\begin{document}

\def\a{\alpha}
\def\b{\beta}
\def\c{\varepsilon}
\def\d{\delta}
\def\e{\epsilon}
\def\f{\phi}
\def\g{\gamma}
\def\h{\theta}
\def\k{\kappa}
\def\l{\lambda}
\def\m{\mu}
\def\n{\nu}
\def\p{\psi}
\def\q{\partial}
\def\r{\rho}
\def\s{\sigma}
\def\t{\tau}
\def\u{\upsilon}
\def\v{\varphi}
\def\w{\omega}
\def\x{\xi}
\def\y{\eta}
\def\z{\zeta}
\def\D{\Delta}
\def\G{\Gamma}
\def\H{\Theta}
\def\L{\Lambda}
\def\F{\Phi}
\def\P{\Psi}
\def\S{\Sigma}

\def\o{\over}
\def\beq{\begin{eqnarray}}
\def\eeq{\end{eqnarray}}
\newcommand{\gsim}{ \mathop{}_{\textstyle \sim}^{\textstyle >} }
\newcommand{\lsim}{ \mathop{}_{\textstyle \sim}^{\textstyle <} }
\newcommand{\vev}[1]{ \left\langle {#1} \right\rangle }
\newcommand{\bra}[1]{ \langle {#1} | }
\newcommand{\ket}[1]{ | {#1} \rangle }
\newcommand{\EV}{ {\rm eV} }
\newcommand{\KEV}{ {\rm keV} }
\newcommand{\MEV}{ {\rm MeV} }
\newcommand{\GEV}{ {\rm GeV} }
\newcommand{\TEV}{ {\rm TeV} }
\def\diag{\mathop{\rm diag}\nolimits}
\def\Spin{\mathop{\rm Spin}}
\def\SO{\mathop{\rm SO}}
\def\O{\mathop{\rm O}}
\def\SU{\mathop{\rm SU}}
\def\U{\mathop{\rm U}}
\def\Sp{\mathop{\rm Sp}}
\def\SL{\mathop{\rm SL}}
\def\tr{\mathop{\rm tr}}

\def\IJMP{Int.~J.~Mod.~Phys. }
\def\MPL{Mod.~Phys.~Lett. }
\def\NP{Nucl.~Phys. }
\def\PL{Phys.~Lett. }
\def\PR{Phys.~Rev. }
\def\PRL{Phys.~Rev.~Lett. }
\def\PTP{Prog.~Theor.~Phys. }
\def\ZP{Z.~Phys. }

\newcommand{\beqr}{\begin{array}}  
\newcommand {\eeqr}{\end{array}}
\newcommand{\la}{\left\langle}  
\newcommand{\ra}{\right\rangle}
\newcommand{\non}{\nonumber}  
\newcommand{\ds}{\displaystyle}
\newcommand{\red}{\textcolor{red}}
\def\ubl{U(1)$_{\rm B-L}$}
\def\REF#1{(\ref{#1})}
\def\lrf#1#2{ \left(\frac{#1}{#2}\right)}
\def\lrfp#1#2#3{ \left(\frac{#1}{#2} \right)^{#3}}
\def\OG#1{ {\cal O}(#1){\rm\,GeV}}

\def\J{{\cal J}}
\def\aa{{\dot \a}}
\def\bb{{\dot \b}}
\def\ss{{\bar \s}}
\def\hh{{\bar \h}}
\def\Z{{\cal Z}}

\baselineskip 0.7cm

\begin{titlepage}

\begin{flushright}
UT-10-04\\
IPMU 10-0056
\end{flushright}

\vskip 1.35cm
\begin{center}
{\large \bf
Notes on Operator Equations of Supercurrent Multiplets and the Anomaly Puzzle in Supersymmetric Field Theories
}
\vskip 1.2cm
Kazuya Yonekura
\vskip 0.4cm

{\it $^1$ Institute for the Physics and Mathematics of 
the Universe (IPMU),\\ 
University of Tokyo, Chiba 277-8568, Japan\\
$^2$  Department of Physics, University of Tokyo,\\
    Tokyo 113-0033, Japan}

\vskip 1.5cm

\abstract{ 
Recently, Komargodski and Seiberg have proposed a new type of supercurrent multiplet which contains the energy-momentum tensor and the supersymmetry current
consistently. In this paper we study quantum properties of the supercurrent in renormalizable field theories. 
We point out that the new supercurrent gives a quite simple resolution to the 
classic problem, called the anomaly puzzle, that the Adler-Bardeen theorem applied to an R-symmetry current is inconsistent with all order corrections to
$\beta$ functions. We propose an operator equation for the supercurrent in all orders of perturbation theory, and then perform several consistency checks of the 
equation. The operator equation we propose is consisitent with the one proposed by Shifman and Vainshtein,
if we take some care in interpreting the meaning of non-conserved currents.
}
\end{center}
\end{titlepage}

\setcounter{page}{2}

\section{Introduction}
\label{sec:1}
In supersymmetric (SUSY) theories, R-charge $R$, supercharge $Q_\a$ and momentum $P^\mu$ 
form a nontrivial algebra $[R,Q_\a] \sim Q_\a $ and $\{Q_\a,\bar{Q}_{\dot \a}\} \sim \s^\m_{\a{\dot \a}}P_\mu$.
This algebra suggests that the corresponding currents $j^\mu_R$, $S^\mu_\a$ and $T^{\mu\nu}$, respectively, are in the same multiplet, 
because the above algebra implies $[j^\mu_R,Q_\a] \sim S^\mu_\a$ and $[S^\mu_\a,\bar{Q}_{\dot \a}] \sim \s^\n_{\a{\dot \a}}T^\m_{\n}$.
In the language of superspace, the multiplet may be represented as a real vector superfield $\J^\mu(x,\h,{\bar \h})$
with 
\beq
\J^\mu(x,\h,{\bar \h}) \sim j^\mu_R+\h S^\mu+{\rm h.c.}+2\h\s_{\n}{\bar \h}T^{\n\m}+\cdots.
\eeq

The existence of such a multiplet, called supercurrent multiplet, was indeed discovered by Ferrara and Zumino~\cite{Ferrara:1974pz}.
The supercurrent of Ferrara and Zumino contains, as its lowest component, the R-symmetry current with the charge assignment $\frac{2}{3}$
to (the lowest components of) 
all chiral matter fields. Later, it was discussed that the anomalies of the R-symmetry current and the trace of the energy-momentum tensor (i.e. the anomaly of 
dilatation transformation)
are also in the same chiral multiplet~\cite{Clark:1978jx,Piguet:1981mu}. The anomaly equation in superspace is given by
\beq
{\bar D}^{\aa} \J_{\a \aa}=D_\a X,
\eeq
where $\J_{\a \aa}=-2\s^\mu_{\a\aa}\J_\mu$ and $X$ is a chiral superfield. Solving this equation in terms of component fields gives
\beq
\frac{2}{3}T^\m_\mu+i\q_\mu j_R^\mu=F_X, \label{eq:anomalyFterm}
\eeq
where $F_X$ is the $F$-term of the chiral field $X$.

As an example, let us consider a SUSY gauge theory with matter fields in the representation $r$ and with no superpotential.
It is known that the trace of the enegy-momentum tensor is proportional to the $\b$ function~\footnote{Our definition of the $\b$ function is given by 
$\b=\q g^2/\q \log\mu$, where $\mu$ is a renormalization scale and $g$ is the gauge coupling.},
\beq
T^\mu_\mu=-\frac{\b(g^2)}{4g^4}F^{\mu\nu}F_{\mu\nu}+\cdots, \label{eq:N0trace}
\eeq
where $F_{\m\n}$ is the field strength of the gauge field, and the sum over the adjoint index is understood in the equation.
We have only shown the term proportional to $F^{\mu\nu}F_{\mu\nu}$ and neglected other terms for simplicity.
On the other hand, because the matter chiral fields have R-charge $\frac{2}{3}$, their fermionic components have R-charge $-\frac{1}{3}$,
and hence the anomaly of the R-symmetry current is given by
\beq
\q_\mu j_R^\mu=\frac{t(A)-\frac{1}{3}t(r)}{16\pi^2} F^{\mu\nu} \tilde{F}_{\mu\nu}+\cdots,\label{eq:ranom}
\eeq
where $t(A)$ and $t(r)$ are the dynkin indices of the adjoint representation and the representation $r$, respectively, and
$\tilde{F}_{\mu\nu}=\frac{1}{2}\e_{\m\n\r\s}F^{\r\s}$. The presence of terms represented by dots will be discussed in the next section, 
but for the time being we neglect it for simplicity.

At the one-loop level, Eq.~(\ref{eq:anomalyFterm}) gives a successful result.
Substituting the one-loop $\b$ function, $\b^{\rm 1-loop}/g^4=-(3t(A)-t(r))/8\pi^2$, we obtain
\beq
\left(\frac{2}{3}T^\m_\mu+i\q_\mu j_R^\mu \right)_{\rm 1-loop} &=& \frac{t(A)-\frac{1}{3}t(r)}{16\pi^2}\left( F^{\mu\nu}F_{\mu\nu}+iF^{\mu\nu} \tilde{F}_{\mu\nu} \right)+\cdots \nonumber \\
&=& -\frac{t(A)-\frac{1}{3}t(r)}{8\pi^2}W^\a W_\a |_{\h^2},
\eeq
where $W^\a$ is the gauge field strength chiral field. Thus, setting 
\beq
X=-\frac{t(A)-\frac{1}{3}t(r)}{8\pi^2}W^\a W_\a,
\eeq
we obtain a consistent result.

However, a problem arises at higher loops.
The $\b$ function, and hence the trace anomaly in Eq.~(\ref{eq:N0trace}), receive corrections in all orders of perturbation theory.
On the other hand, the Adler-Bardeen theorem~\cite{Adler:1969er} asserts that the anomaly of the R-symmetry current is exhausted at the one-loop level,
and there are no higher order corrections to Eq.~(\ref{eq:ranom}). However, Eq.~(\ref{eq:anomalyFterm}) requires that they must be combined to form a
$F$-term of some chiral field. This dilemma is called the anomaly puzzle. 

There have been several attempts to resolve the anomaly puzzle,  
and we believe that some of them~\cite{Grisaru:1985yk,Shifman:1986zi} are successful and 
indeed give the basis of the present work (see also Ref.~\cite{Huang:2010tn} for a recent discussion).
However, they are technically quite involved, and some points in their discussions seem to be not yet clear 
(at least to the authour, which we will discuss in the later sections).
There is also a work~\cite{ArkaniHamed:1997mj} which gives a intuitively clear argument, but no explicit equation of the supercurrent 
was written down in that work.

The origin of the problem is that the supercurrent must satisfy Eq.~(\ref{eq:anomalyFterm}) with some chiral field $X$, which gives a very stringent constraint 
on the supercurrent.
However, Komargodski and Seiberg have recently shown~\cite{Komargodski:2010rb} 
that the energy-momentum tensor and the supersymmetry current can be in a same supercurrent 
multiplet even if Eq.~(\ref{eq:anomalyFterm}) is not satisfied (see also Ref.~\cite{Magro:2001aj} for early work). 
They have shown that it is sufficient (although there are other possibilities~\cite{Kuzenko:2010am}) for a supercurrent to satisfy the equation~\footnote{
We use $\J_\m$ for the supercurrent regardless of the equation it satisfies. Also, we have represented the $\chi_\a$ of Ref.~\cite{Komargodski:2010rb} as $\chi_\a=-\frac{1}{4}D^2 \bar{D}_{\aa} { J}$, which gives rise to no 
problem in the present paper. }
\beq
{\bar D}^{\aa} \J_{\a \aa}=D_\a X-\frac{1}{4}\bar{D}^2 D_{\a}{J},
\eeq
for a chiral field $X$ and a real vector field ${J}$. 

Solving the above equation gives~\cite{Komargodski:2010rb}
\beq
\frac{2}{3}T^\m_\m &=& {\rm Re}F_X-\frac{1}{6}D_{J}, \label{eq:traceanom} \\
\q_\m j_R^\m &=& {\rm Im}F_X, \label{eq:Ranom}
\eeq
where $D_{ J}$ is the $D$-term of ${ J}$. Now, the trace anomaly is not equal to the real part of the $F$-term of the chiral field $X$ !
This suggests a quite simple resolution to the anomaly puzzle; if $X$ is exhausted at the one-loop level and $ J$ receives higher order corrections,
there is in fact no conflict between the Adler-Bardeen theorem and the all order nature of the trace anomaly.

This paper is organized as follows. In section~2, we propose a supercurrent equation which is supposed to be valid to all orders of perturbation theory,
based on the above resolution to the anomaly puzzle. 
We also discuss the relation of our proposal to the works Refs.~\cite{Grisaru:1985yk}, \cite{Shifman:1986zi}.
Section~3 gives consistency checks on the proposed equation. 
Section~4 studies the trace anomaly Eq.~(\ref{eq:traceanom}) in Wilson effective field theory along the line of Refs.~\cite{Shifman:1986zi,ArkaniHamed:1997mj}.
\ref{app:A} collects our notation and conventions.
\ref{app:B} contains the discussion of the definition and properties of non-conserved currents, which is necessary to clarify the meaning of 
current operator equations.

\section{The operator equation of supercurrent}
\label{sec:2}
In this section, we propose an operator equation which is supposed to be valid to all orders of perturbation theory
(and even non-perturbatively in many cases).
We consider a theory with chiral fields $\F_r$ in the representations $r$ of some gauge group $G$. The chiral fields 
also have a renormalizable superpotential $W(\F)$ and a canonical K\"ahler potential.
We take the R-symmetry current $j_R^\mu$ with charge assignment $\frac{2}{3}$ to all chiral matter fields,
and we define the supercurrent $\J^\m$ to have this R-symmetry current as its lowest component. 
Then, we propose the following supercurrent equation; if operators are defined and renormalized appropriately at the quantum level, 
the supercurrent equation is given as
\beq
{\bar D}^{\aa} \J_{\a \aa}=D_\a X-\frac{1}{4}\bar{D}^2 D_{\a}{J}, \label{eq:supercurrenteq}
\eeq
 where $X$ and ${J}$ are given by
 \beq
 X&=& \frac{4}{3}\left[3W-\sum_r \F_r\frac{\q W}{\q \F_r} -\frac{3t(A)-\sum_r t(r)}{32\pi^2}W^\a W_\a \right], \label{eq:anompart} \\
 {J}&=&-2\sum_{r}\g_r \F_r^\dagger e^{-2V} \F_r . \label{eq:correctionpart}
 \eeq
$\g_r$ is the anomalous dimension~\footnote{
The definition of the anomalous dimensions and composite operators
depends on renormalization procedures (and gauge fixings for gauge non-invariant operaters). Thus, Eq.~(\ref{eq:supercurrenteq}) is valid 
only for ``appropriate'' renormalization procedures. 
Throughout this paper we simply assume the existence of such renormalization procedures, which satisfy several desired properties such as 
the NSVZ $\b$ function discussed later and the Adler-Bardeen theorem.
Various discussions found in the literature and in this paper motivate the existence of such
``appropriate'' renormalization procedures.  
} of $\F_r$. We have assumed a diagonal anomalous dimension matrix for chiral fields for simplicity, but a generalization is straightforward.

This supercurrent equation is determined as follows.
In the chiral multiplet $X$, there are terms which depend on the superpotential. These terms represent the tree-level violation of the conservation of $j_R^\mu$
(see Eq.~(\ref{eq:Ranom})).
The term proportional to $W^\a W_\a$ is the one-loop anomaly as discussed in the Introduction. Assuming the Adler-Bardeen theorem (although the theorem has
some ambiguity discussed in \ref{app:B}), we insist that $X$ be exhausted at the one-loop level. Then, $X$ should be exactly given by Eq.~(\ref{eq:anompart}).
Higher order corrections are represented by $J$. From a simple dimensional analysis and gauge invariance, we can see that $J$ should be a linear combination of $\F^\dagger e^{-2V}\F$.
(Note that terms like $\F\F+{\rm h.c.}$ give no contribution in $\bar{D}^2D_\a J$.) The coefficients of these operators are 
chosen so as to be consistent with the later discussions in this and subsequent sections. 

Following the discussion of Ref.~\cite{Komargodski:2010rb}, we can always define another supercurrent by using a real vector superfield $U$,
\beq
\J_{\a\aa} &\to& \J_{\a\aa}'=\J_{\a\aa}+[D_\a,\bar{D}_\aa]U, \\
X &\to& X'=X+\frac{1}{2}\bar{D}^2U, \\
J &\to & J'=J-6U.
\eeq
In components, this redefinition of the supercurrent gives rise to the redefinition of the R-symmetry current $j_R^\m$, the supersymmetry current $S^\mu_\a$ and 
the energy-momentum tensor $T^{\mu\nu}$. Using the equations collected in \ref{app:A}, we have~\cite{Komargodski:2010rb}
\beq
j_R^\m &\to& j_R^\m-j_U^\m, \label{eq:Rredefine}\\
S^\m &\to& S^\m+\frac{1}{2}(\s^\m \ss^\n-\s^\n \ss^\m)\q_\n \chi_U, \\
T_{\m\n} &\to& T_{\m\n}+\frac{1}{2}(\q_\m\q_\n-\y_{\m\n}\q^2)C_U, \label{eq:improvement}
\eeq
where $j_U^\m$ is the $\h\s^\m\hh$ component of $U$, $\chi_U$ the $\h$ component of $U$ and $C_U$ the lowest component of $U$.
The changes of $T_{\m\n}$ and $S^\m_\a$ are improvements of these currents, but $j_R^\m$ becomes a totally different R-current.

In particular, if we choose $6U=J$, we have a new supercurrent $\J_{\m}^{\rm FZ}$
\beq
{\bar D}^{\aa} \J^{\rm FZ}_{\a \aa}=D_\a X^{\rm FZ} \label{eq:FZequation},
\eeq
where 
\beq
X^{\rm FZ}=\frac{4}{3}\left[3W-\sum_r \F_r\frac{\q W}{\q \F_r} -\frac{3t(A)-\sum_r t(r)}{32\pi^2}W^\a W_\a -\frac{1}{8}\sum_{r} \g_r \bar{D}^2 (\F_r^\dagger e^{-2V} \F_r)\right]. \label{eq:shifvain}
\eeq
$\J^{\rm FZ}_{\a\aa}$ is now the ``Ferrara-Zumino multiplet'' in the sense that it satisfies Eq.~(\ref{eq:FZequation}).
Furthermore, by using the Konishi anomaly equation~\cite{Konishi:1983hf} (which is the extension of an ordinary anomay equation to superspace)
\beq
\frac{1}{4}\bar{D}^2 (\F_r^\dagger e^{-2V}\F_r)=\F_r\frac{\q W}{\q \F_r}+\frac{t(r)}{16\pi^2}W^\a W_\a, \label{eq:konishiequation}
\eeq
we obtain
\beq
X^{\rm FZ}=\frac{4}{3}\left[3W-\sum_r \left(1+\frac{\g_r}{2}\right)\F_r\frac{\q W}{\q \F_r} -\frac{3t(A)-\sum_r (1-\g_r)t(r)}{32\pi^2}W^\a W_\a \right].\label{eq:modifiedanom}
\eeq
Note that the coefficient of $W^\a W_\a$ is the numerator of the Novikov-Shifman-Vainshtein-Zakharov (NSVZ) 
exact $\b$ function~\cite{Novikov:1983uc,Shifman:1986zi,ArkaniHamed:1997mj}, 
\beq
\b^{\rm NSVZ}=-\frac{g^4}{8\pi^2}\frac{3t(A)-\sum_r(1-\g_r)t(r)}{1-t(A)g^2/8\pi^2}, \label{eq:NSVZ}
\eeq 
which motivates the form of $J$ in Eq.~(\ref{eq:correctionpart}).

Now let us comment on the previous solutions to the anomaly puzzle found in the literature~\cite{Grisaru:1985yk,Shifman:1986zi}, in light of the new supercurrent 
equation of Ref.~\cite{Komargodski:2010rb}. In Ref.~\cite{Grisaru:1985yk}, it was discussed that one can define two distinct supercurrents.
The lowest component of one supercurrent satisfies the Adler-Bardeen theorem, but it does not contain the conserved energy-momentum tensor.
The other supercurrent contains the energy-momentum tensor, but it does not satisfy the Adler-Bardeen theorem.
The authors of Ref.~\cite{Grisaru:1985yk} used regularization by dimensional reduction (DRED)~\cite{Siegel:1979wq,Capper:1979ns}, and discussed that 
the difference of the two supercurrents arises from the ambiguity of the definition of the supercurrents in $\e$ dependent terms, where $\e=4-d$
is the dimensions which is dimensionally reduced in the regularization procedure. 
Now, at least as to the chiral matter contributions, we can understand their result quite simply. 
We have two supercurrents satisfying Eqs.~(\ref{eq:supercurrenteq},\ref{eq:anompart},\ref{eq:correctionpart}) 
and Eqs.~(\ref{eq:FZequation},\ref{eq:modifiedanom}). $\J_\m$ in Eq.~(\ref{eq:supercurrenteq}) satisfies the Adler-Bardeen theorem,
and $\J^{\rm FZ}_\mu$ in Eq.~(\ref{eq:FZequation}) satisfies the Ferrara-Zumino equation. These two supercurrents correspond to the ones considered in
Ref.~\cite{Grisaru:1985yk}.
However, the current $\J_\m$ also contains the conserved energy-momentum tensor as discovered in Ref.~\cite{Komargodski:2010rb},
and the difference between $\J_\m$ and $\J_\m^{\rm FZ}$ can be understood in terms of fully renormalized operators, without any reference
to $\e$ dimensions. 

One point which is still mysterious is the contribution to the $\b$ function in the denominator of the NSVZ $\b$ function.
For example, in a pure ${\cal N}=1$ Yang-Mills theory, the $\b$ function still receives all order corrections due to the denominator in Eq.~(\ref{eq:NSVZ}).
The existence of two supercurrents was discussed in Ref.~\cite{Grisaru:1985yk} even in this case, while we cannot have two supercurrents in the discussion of the 
present section without using DRED. If one uses DRED, the equation which appeared in Ref.~\cite{Grisaru:1985yk}
in the case of the pure Yang-Mills is of the form Eq.~(\ref{eq:supercurrenteq}) with
\beq
X &=&-\frac{4}{3}\left[\frac{3t(A)}{32\pi^2}W^\a W_\a \right] \nonumber \\
J &\propto& g_\e^{\m\n}\G_\m \G_\m,\label{eq:evanescent}
\eeq
where $\G_\m$ is the superfield extension of the gauge connections~\cite{Grisaru:1985tc}, and $g_\e^{\m\n}$ is the metric in the $\e$ ``compactified'' dimensions.
Because the gauge connections in the $\e$ dimensions are only adjoint fields, the right hand side of Eq.~(\ref{eq:evanescent}) is gauge invariant.
Furthermore, $g_\e^{\m\n}\G_\m \G_\m$ satisfies
\beq
\bar{D}^2\left(g_\e^{\m\n}\G_\m \G_\m\right) \propto \e W^\a W_\a.
\eeq
Thus, by a supercurrent redefinition $\J_{\a\aa}+[D_a, \bar{D}_\aa]U$, we may obtain Eq.~(\ref{eq:FZequation}) with
\beq
X^{\rm FZ}=-\frac{4}{3}\left[\left(\frac{3t(A)}{32\pi^2}+{\rm higher~corrections?}\right)W^\a W_\a \right]. \label{eq:nazo}
\eeq
It may be that the full NSVZ $\b$ function (i.e.~including the denominator) appears in Eq.~(\ref{eq:nazo}). However, even if so,
the above mechanism strongly depends on DRED, with no obvious regularization-independent argument. Furthermore, it is difficult to
understand why the effect of $\e$ dimensions remains after taking the limit $\e \to 0$ if the operators are appropriately renormalized 
(i.e.~all $1/\e$ poles are appropriately subtracted).
Perhaps more detailed diagrammatic study is required to see the consistency, which we leave for future work. Instead, in Section~\ref{sec:3} we perform 
consistency checks on Eqs.~(\ref{eq:supercurrenteq},\ref{eq:anompart},\ref{eq:correctionpart}) based on exact results established in SUSY theories,
and in Section~\ref{sec:4} we derive Eq.~(\ref{eq:traceanom}) by a formal argument in Wilsonian effective field theory. 
All of the discussions of those sections suggest that the 
denominator of the NSVZ $\b$ function does not (and should not) appear in the supercurrent equation.
Thus, we do not claim that Eq.~(\ref{eq:nazo}) is a solution to the above problem.

In fact, it is not clear (at least to the author) whether the non-appearance of the denominator of the NSVZ $\b$ function in the supercurrent 
equation is really a problem or not.
In general, the definition of couplings and $\b$ functions depend on renormalization schemes
(e.g. DR, $\overline{\rm DR}$, subtraction at a Euclidean momentum point, etc.). The definition of composite operators is also 
renormalization scheme dependent. Therefore, the trace anomaly equation (\ref{eq:N0trace}) should be valid only for 
some combinations of renormalization schemes for the gauge coupling and the operator $F^{\mu\nu}F_{\mu\nu}$,
and if we use other renormalization procedures, there seems to be no reason why Eq.~(\ref{eq:N0trace}) should precisely be satisfied.
However, there is one important motivation for the trace anomaly to be proportional to $\b$ functions.
It is widely believed that when a theory is scale invariant, then the theory is also conformal invariant~\footnote{In some non-unitary Euclidean field theories,
there are known counterexamples to the equivalence of scale and conformal invariance. We would like to thank Y.~Nakayama for pointing out this
to us. }. 
Thus, if $\b$ functions vanish (i.e. the theory is scale invariant), then it is expected that the trace anomaly also vanishes (i.e. the theory is conformal invariant).
For this to be true, it is enough for the trace anomaly to be proportional to the numerator of the NSVZ $\b$ function.
The denominator of the NSVZ $\b$ function cannot provide a zero point of the $\b$ function. It can only provide a pole, but the pole cannot be reached
at least when perturbation is valid. On the other hand, the numerator of the NSVZ $\b$ function can become zero even in 
weakly coupled theories~\cite{Banks:1981nn,Seiberg:1994pq}
and hence the trace anomaly should be proportional to it, or else the equivalence of scale and conformal invariance is invalidated.

Shifman and Vainshtein proposed~\cite{Shifman:1986zi} the operator equation Eqs.~(\ref{eq:FZequation},\ref{eq:shifvain})
with the lowest component of $\J^{\rm FZ}$ given by the R-symmetry current with charge assignment $\frac{2}{3}$ to all chiral matter fields.
At first glance, this contradicts with our supercurrent equation, since we must redefine the R-symmetry current as in Eq.~(\ref{eq:Rredefine}) to
obtain Eqs.~(\ref{eq:FZequation},\ref{eq:shifvain}), and hence the charge assignment of the R-symmetry seems to be changed. 
In fact, the $\h\s^\m\hh$ component of the Konishi current $J_r \equiv \F_r^\dagger e^{-2V}\F_r$, which we denote $j^\m_{J_r}$,
 is an anomalous current with charge assignment $+1$ to the chiral field $\F_r$, and zero to other fields. The redefinition to obtain Eqs.~(\ref{eq:FZequation},\ref{eq:shifvain}) leads
 \beq
 j_R^\m \to j_R^{\rm (FZ)\m}= j_R^\m+\sum_r \frac{1}{3}\g_r j^\m_{J_r},\label{eq:newRcurrent}
 \eeq
giving another R-symmetry current. 

We can interpret the supercurrent equation of Ref.~\cite{Shifman:1986zi} as follows. 
The new R-symmetry current $j_R^{\rm (FZ)\m}$ in Eq.~(\ref{eq:newRcurrent}) is indeed an R-symmetry current with charge assignment
$\frac{2}{3}$ to all chiral fields, with anomaly given by ${\rm Im} F_{X^{\rm FZ}}$ of Eq.~(\ref{eq:shifvain}), {\it but not} Eq.~(\ref{eq:modifiedanom}).
When we obtain Eq.~(\ref{eq:modifiedanom}) from Eq.~(\ref{eq:shifvain}), we use the anomaly equation Eq.~(\ref{eq:konishiequation}).
However, when we consider Ward identities involving the currents, the anomaly equation is valid only up to contact terms, and the contact terms are 
essencial for the definition of the currents. In fact, contact terms are the ones which determine charges of fields.
See \ref{app:B} for details.
Thus, both $j_R^\m$ and $j_R^{\rm (FZ)\m}$ are currents with charge assignment $\frac{2}{3}$ to all chiral fields, if we take care that 
the anomaly of $j_R^{\rm (FZ)\m}$ is given by Eq.~(\ref{eq:shifvain}) instead of Eq.~(\ref{eq:modifiedanom}).
With this interpretation, our supercurrent equation is consistent with that of Ref.~\cite{Shifman:1986zi}.

Before closing this section, let us discuss one more subtle point regarding the anomaly puzzle.
Neglecting the superpotential, the anomaly of the R-current is proportional to the imaginary part of the $F$-term of $W^\a W_\a$,
\beq
{\rm Im} \left( W^\a W_\a|_{\h^2} \right) = -\frac{1}{2}F^{\m\n}\tilde{F}_{\m\n}-\q_\mu (\l \s^\mu \bar{\l}), \label{eq:gauginodousiyou}
\eeq
where $\l$ is a gaugino.
It contains a contribution from the gaugino other than the gauge field.
The usual Adler-Bardeen theorem in ${\cal N}=0$ theories is supposed to state that the anomaly should be given by one-loop exact expression 
proportional to $F^{\m\n}\tilde{F}_{\m\n}$, without any other terms. On the other hand, ${\cal N}=1$ supermultiplet structure dictates that we should include the gaugino term. However, ${\cal N}=1$ theories are merely a subset of ${\cal N}=0$ theories.
How should we consider about the gaugino contribution?

It is often said that the one-loop exactness of the anomaly is related to the topological nature of $F^{\m\n}\tilde{F}_{\m\n}$.
In instanton backgrounds, the integral of $F^{\m\n}\tilde{F}_{\m\n}$ times some factor is quantized to be integers, and this integer can be interpreted as violation of
charge conservation of anomalous symmetries by instantons~\cite{'tHooft:1976up}.
The point is that this property is {\it not} violated by the presence of the gaugino term in Eq.~(\ref{eq:gauginodousiyou}).
This is because the term $\q_\mu (\l \s^\mu \bar{\l})$ is a total derivative of a gauge invariant operator, and the integration of it simply vanishes.
Thus the integration of Eq.~(\ref{eq:gauginodousiyou}) is the same as the integration of $-\frac{1}{2}F^{\m\n}\tilde{F}_{\m\n}$, which is quantized.

Let us see more explicitly how the integration of $\q_\mu (\l \s^\mu \bar{\l})$ vanishes.
Consider a Euclidean correlation function $\vev{\l \s^\mu \bar{\l}(y){\cal O}_1(x_1){\cal O}_1(x_2)\cdots}$ in some fixed instanton background, and take
the limit $y \to \infty$. The propagator of gaugino behaves as ${\cal O}(r^{-3})$, where $r$ is the distance between two gauginos.
(Here we considered the propergator of gauginos where one gaugino is at the position $y$ and the other at some $x_i$.)
The gaugino zero modes in instanton background behave as ${\cal O}(r^{-3})$ or ${\cal O}(r^{-4})$, 
depending on zero modes  (see e.g. Ref.~\cite{Shifman:1999mv}), where $r$ is the distance
between an instanton and a gaugino. (Here the gaugino is at $y$ and the instanton is at some fixed position.)
Regardless of whether the gauginos in the operator $\l \s^\mu \bar{\l}$ are contracted with other gauginos or with instantons,
the above correlation function behaves at most as 
\beq
\vev{\l \s^\mu \bar{\l}(y){\cal O}_1(x_1){\cal O}_1(x_2)\cdots} \to {\cal O}((y^{-3})^2)~~~~(y \to \infty),
\eeq
where we have taken into account that the operator $\l \s^\mu \bar{\l}$ contains two gauginos.
Thus, if $\l \s^\mu \bar{\l}$ is integrated on an infinite sphere, it behaves as ${\cal O}((y^{-3})^2 \cdot y^3) \to 0$. Therefore, the
integration of the total derivative $\q_\mu (\l \s^\mu \bar{\l})$ gives no contribution.

It may seem puzzling that the gaugino term gives no contribution by the following reason. 
Let us consider the case of ${\cal N}=1$ pure Yang-Mills.
Then, the operator $\l \s^\mu \bar{\l}$ is proportional to the R-symmetry current $j_R^\mu$, and hence it satisfies the anomaly equation
$\q_\mu (\l \s^\mu \bar{\l}) \sim g^2F^{\m\n}\tilde{F}_{\m\n}$. We have discussed that the integration of the left hand side of this equation vanishes,
while the integration of the right hand side is obviously nonzero in instanton backgrounds. 
The resolution to this puzzle comes from contact terms. As discussed in \ref{app:B}, the anomaly equation is valid only up to contact terms,
and the integration of $g^2F^{\m\n}\tilde{F}_{\m\n}$ should be balanced with the integration of the contact terms. Then there is no puzzle that 
the integration of $\q_\mu (\l \s^\mu \bar{\l})$ is zero.

In conclusion, we claim that the presence of the gaugino term in Eq.~(\ref{eq:gauginodousiyou}) does not affect the topological argument given above,
and hence it does not spoil the ``Adler-Bardeen theorem'' in some sense. See \ref{app:B} for further details.

\section{Consistency checks}
\label{sec:3}
In this section we perform consistency checks on the proposals of the previous section.
See also Ref.~\cite{Shifman:1999mv}
and references therein.

\subsection{Conformal fixed point in ${\cal N}=1$ gauge theories}\label{sec:3.1}
In many SUSY gauge theories, there exists a conformal fixed point~\cite{Seiberg:1994pq}. Some of the properties of the fixed point
is exactly known due to superconformal symmetry~\cite{Flato:1983te}. We check the consistency of 
Eqs.~(\ref{eq:supercurrenteq},\ref{eq:anompart},\ref{eq:correctionpart}) with those results.

At the conformal fixed point, all the anomalous dimensions are constants (i.e. independent of the renormalization scale).
The supercurrent $\J_\m$ in Eq.~(\ref{eq:supercurrenteq}) contains the R-symmetry current with charge assignment $\frac{2}{3}$ to
all chiral fields. Then, by using the redefinition of the current as in Eq.~(\ref{eq:Rredefine}) and using the Konishi anomaly equation Eq.~(\ref{eq:konishiequation}),
we obtain Eqs.~(\ref{eq:FZequation},\ref{eq:modifiedanom}). As discussed in the previous section and in \ref{app:B}, the use of the anomaly equation
changes the charge assignment of the R-current. From Eq.~(\ref{eq:newRcurrent}), one can see that the new R-symmetry assigns charge 
\beq
\frac{2}{3}+\frac{\g_r}{3} \label{eq:fixedpointrcharge}
\eeq
to (the lowest component of) chiral fields $\F_r$.

The $\b$ function Eq.~(\ref{eq:NSVZ}) should vanish at the fixed point, and hence the term proportional to $W^\a W_\a$ in Eq.~(\ref{eq:modifiedanom})
vanishes. Note that the denominator of the NSVZ $\b$ function plays no role in this discussion.
As to the superpotential terms, if the superpotential contains a term of the form $W \sim \F_{1}\F_{2}\cdots \F_{m}$, then
\beq
3W-\sum_r \left(1+\frac{\g_r}{2}\right)\F_r\frac{\q W}{\q \F_r}  \sim \left(3-m-\frac{1}{2}\sum_i\g_i\right)\F_{1}\F_{2}\cdots \F_{m}.\label{eq:supermarginal}
\eeq
The condition that the parenthesis in the right hand side of Eq.~(\ref{eq:supermarginal}) vanishes is just the condition that the 
interaction $W \sim \F_{1}\F_{2}\cdots \F_{m}$ is a margial interaction (i.e. not relevant or irrelevant interaction) at the fixed point.
Thus, Eq.~(\ref{eq:modifiedanom}) completely vanishes and the new supercurrent (which we call $\J^{\rm SC}_\m$) satisfies the equation~\cite{Komargodski:2010rb} 
\beq
\bar{D}^\aa \J^{\rm SC}_{\a\aa}=0.
\eeq
This equation in particular implies the conservation of the R-symmetry current and the tracelessness of the energy-momentum tensor.

At the fixed point, a scaling dimension $\D$ and an anomalous dimension $\g$ of a scalar field satisfy the relation $\D=1+\g/2$. This relation can be obtained
by solving the two point correlation function of the field in terms of the anomalous dimension by using the Callan-Symanzik equation, and then compare the
result to the exact expression for the correlation function in conformal field theory. Thus, the charge Eq.~(\ref{eq:fixedpointrcharge}) implies
the relation between the R-charge $R_r$ and the scaling dimension $\D_r$ of the (lowest component of) chiral field $\F_r$,
\beq
R_r=\frac{2}{3}\D_r.\label{eq:chiralprimaryeq}
\eeq
It is known that the ${\cal N}=1$ superconformal algebra requires the existence of the R symmetry, and the R-charge and the scaling
dimension of chiral (primary) fields must satisfy 
exactly the relation Eq.~(\ref{eq:chiralprimaryeq})~\footnote{Strictly speaking, what is rigorously proved is that Eq.~(\ref{eq:chiralprimaryeq}) is satisfied by
a gauge invariant chiral primary field which may or may not be a composite operator. But practically, Eq.~(\ref{eq:chiralprimaryeq})
is valid even for gauge non-invariant fields which is not a composite operator.}.
Thus, the R-symmetry in $\J^{\rm SC}_\m$ is precisely the one which appears in the superconformal algebra, giving consistent result.

\subsection{Central extension of ${\cal N}=1$ algebra and BPS domain walls}
${\cal N}=1$ SUSY algebra has a central extension with applications to strong dynamics of SUSY gauge theories~\cite{Dvali:1996xe}.
By calculating the supersymmetric transformation of the supercurrent in Eq.~(\ref{eq:supercurrenteq}), one obtains~\cite{Komargodski:2010rb} 
(see \ref{app:A} for notation)
\beq
\{S_{\m\a},\bar{Q}_\bb\}&=&\s^\n_{\a\bb}\left( 2T_{\m\n}+\frac{1}{4}\e_{\n\m\r\s}\q^\r j_J^\s -i\y_{\n\m}\q_\r j_R^\r+i\q_\n j_{R\m}-
\frac{1}{2}\e_{\n\m\r\s}\q^\r j^\s_R \right), \label{eq:algebra1} \\
\{S_{\m\a},Q^\b\}&=&-2i(\s_{\m\n})^\b_\a\q^\n \f_X^\dagger. \label{eq:algebra2}
\eeq
Taking the $\mu=0$ component and integrating over the space $\int d^3x$ gives the SUSY algebra. 
We assume that all fields which have space-time indices vanish first enough at spatial infinity so that the surface terms
do not contribute in the integration. Then, Eq.~(\ref{eq:algebra1}) gives ordinary SUSY algebra. 
However, $\f_X$ need not vanish at spatial infinity because it is the lowest component of the chiral field $X$.
If there exists a domain wall in the theory, the value of $\f_X$ may be different between two phases separated by the wall.
In that case, by integrating Eq.~(\ref{eq:algebra2}) we obtain
\beq
\{Q_\a,Q^\b\} &=&-2i(\s_{0i})^\b_\a n_i A \Z ,\\
\Z&=&\vev{\f_X^\dagger}_2-\vev{\f_X^\dagger}_1,
\eeq
where $A$ is the wall area (which is infinity if the wall is present in flat four dimenional spacetime, but that does not matter in the present discussion),
$n_i$ is the unit space vector orthogonal to the wall, and $\vev{\f_X^\dagger}_a~(a=1,2)$ is the vacuum expectation value of $\f_X^\dagger$
in the two phases separated by the wall. We have assumed that the wall is static.

As is usual in the central extension of the SUSY algebra, the central charge gives a lower bound on the wall energy.
The wall tension is bounded as
\beq
{\rm Wall~tension} \geq \frac{1}{2}|\Z|,
\eeq
with the equality saturated if the wall is in a BPS state. See Ref.~\cite{Chibisov:1997rc} for a detailed discussion on these matters.
Note that the one-loop exactness of $X$ means the one-loop exactness of the central charge $\Z$.

Let us apply the above formulation to the case of ${\cal N}=1$ pure $SU(N)$ Yang-Mills~\cite{Dvali:1996xe}.
In this case, gaugino condensation occurs, with the vacuum expectation value of $W^\a W_\a$ exactly given as
(see Ref.~\cite{Intriligator:1995au} for a review)
\beq
\vev{-\frac{W^\a W_\a}{32\pi^2}}=e^{2\pi i k/N}\L^3~~~~(k=1,2,\cdots,N),
\eeq
where $\L$ is the holomorphic dynamical scale, and $k=1,2,\cdots,N$ represents the existence of $N$ vacua in pure $SU(N)$ theory. 
In this theory we can have a domain wall since there are $N$ phases of vacua.
Then, using Eq.~(\ref{eq:anompart}), $\Z$ is given by
\beq
\frac{1}{2}|\Z|=2N\left| \exp(2\pi i k_2 /N)-\exp(2\pi i k_1/N) \right| |\L|^3, \label{eq:purecentral}
\eeq
where $k_1$ and $k_2$ are the values of $k$ of the two phases separated by the wall.

The central charge is a physical observable as the tension of the BPS domain wall.
Thus, from Eq.~(\ref{eq:purecentral}), we can see that $\L$ should also be a physical observable. 
Renormalization group property of holomorphic dynamical scales is studied in detail in Ref.~\cite{ArkaniHamed:1997ut}.
It is shown that a holomorphic dynamical scale $\L$ is not invariant under renormalization group flow in general gauge theories with matter fields.
However, in the case of pure Yang-Mills, $\L$ is invariant under renormalization group, and hence is a physical observable. 
This is consistent with the appearance of $\L$ in Eq.~(\ref{eq:purecentral}).
If the coefficient of $W^\a W_\a$ in Eq.~(\ref{eq:anompart}) (or equivalently Eq.~(\ref{eq:modifiedanom}) in the case of pure Yang-Mills)
were not the one-loop exact expression but  $\b^{\rm NSVZ}/g^4$ (see Eq.~(\ref{eq:NSVZ})), we would obtain a renormalization group non-invariant
central charge, since the gauge coupling is not renormalization group invariant. This is an evidence that the full NSVZ $\b$ function
should not appear in Eq.~(\ref{eq:modifiedanom}), and only the numerator of the NSVZ $\b$ function appears.

\subsection{One-loop exactness in ${\cal N}=2$ SUSY QCD} \label{sec:3.3}
${\cal N}=2$ gauge theories are special cases of ${\cal N}=1$ gauge theories, so the formulation of the previous section
should also apply to ${\cal N}=2$ theories. Let us consider ${\cal N}=2$ massless SUSY QCD (SQCD). (Inclusion of mass terms is straightforward.)
In ${\cal N}=1$ language, we have vector-like pairs of chiral fields $Q_r,\tilde{Q}_r$ in the representation $r$ and $\bar{r}$ respectively,
which form hypermultiplets, 
and an adjoint field $\F_{Ad}$ which, combined with the gauge multiplet, form an ${\cal N}=2$ vector multiplet.
To coincide with the previous convention, let us canonically normalize $\F_{Ad}$ for the time being. Then the superpotential is given by
\beq
W=\sqrt{2}g\sum_r \tilde{Q}_r \F_{Ad} Q_r.
\eeq
It is known that the anomalous dimensions of $Q_r,\tilde{Q}_r$ are exactly zero (see e.g. Ref.~\cite{Argyres:1996eh} for a non-perturbative argument).
The wave function renormalization of $\F_{Ad}$ is given by the gauge coupling constant because it is in the same multiplet with the gauge field,
and hence its anomalous dimension is given by the $\b$ function as
\beq
\g_{{\rm \F}_{Ad}}=-\m\frac{\q}{\q \m} \log g^{-2}=\frac{\b}{g^2}.
\eeq
It is also known that the $\b$ functions in ${\cal N}=2$ theories are one-loop exact.
Using these anomalous dimensions, $J$ in Eq.~(\ref{eq:correctionpart}) is given by
\beq
J=-2\frac{\b}{g^2}\F_{Ad}^\dagger e^{-2V}\F_{Ad}. \label{eq:JinN2}
\eeq

In Section~\ref{sec:2}, we have stated that $X$ is the one-loop contribution and $J$ represents the higher order corrections. 
If so, Eq.~(\ref{eq:JinN2}) shows that the supercurrent equation in the ${\cal N}=2$ SQCD receives higher order corrections. 
However, this distinction of the one-loop and the higher order contributions is in fact not necessarily well-defined,
and the corrections to the supercurrent equation of ${\cal N}=2$ SQCD is really exhausted at the one-loop level to all orders of perturbation theory
as we will now see. For this purpose, we use the redefinition of the supercurrent as in Section~\ref{sec:2} to obtain
Eqs.~(\ref{eq:FZequation},\ref{eq:shifvain}). Furthermore, we redefine the normalization of $\F_{Ad}$ as 
\beq
\F_{Ad} \to \frac{1}{g}\F_{Ad},
\eeq
so as to be consistent with ${\cal N}=2$ structure.
We also use the one-loop-exact $\b$ function $\b/g^4=-(2t(A)-\sum_r t(r))/8\pi^2$.
Denoting the new supercurrent by $\J_\m^{{\cal N}=2}$, we obtain
\beq
\bar{D}^\aa \J_{\a\aa}^{{\cal N}=2}=D_\a X^{{\cal N}=2},
\eeq
where
\beq
X^{{\cal N}=2}=-\frac{2t(A)-\sum_r t(r)}{24\pi^2}\left(W^\a W_\a-\frac{1}{2} \bar{D}^2 (\F_r^\dagger e^{-2V} \F_r)  \right).
\eeq
Note that in the above redefinition of the supercurrent, the charge assignment of the R-symmetry does not change
because we have not used any anomaly equation, as explained in detail in \ref{app:B}.
Note also that the combination in the parenthesis is exactly the one which appears in the ${\cal N}=2$ gauge kinetic term in the Lagrangian.
Thus, in this form it is clear that the corrections to the supercurrent equation is exhausted at the one-loop level to all orders of perturbation theory,
if the supercurrent is appropriately defined.

\section{Trace anomaly in Wilson effective field theory and holomorphic gauge coupling}\label{sec:4}
In the previous two sections, we have proposed the supercurrent equation and seen the consistency of the equation.
In this section we study the trace anomaly in Wilson effective theory along the line of Refs.~\cite{Shifman:1986zi,ArkaniHamed:1997mj}.
The argument of this section
comes near the proof of Eq.~(\ref{eq:traceanom}) with $F_X$ and $D_J$ given by the $F$-term and $D$-term of Eqs.~(\ref{eq:anompart})
and (\ref{eq:correctionpart}), respectively.

Before discussing the supersymmetric case, let us give a general argument which is applicable even for non-SUSY cases.
We consider a theory described by some fields $\f_i$ with mass dimension $\D^c_i$ (where $c$ in the superscript means ``classical scaling dimension''). 
There are coupling constants $g_a$ in the theory with mass dimension $d_a$, and the theory is regularized with the Wilson cutoff scale $M$.
Among the coupling constants, we include the wave function renormalization factors of the fields so that we do not renormalize the fields as we change the 
cutoff scale $M$. Then, we consider the following correlation function,
\beq
Z(g_a, M, \{i,j,\cdots\}) \equiv \vev{\f_i(x_1) \f_j(x_2)\cdots}.
\eeq
This correlation function $Z$ is calculated by using path integral as
\beq
Z(g_a, M, \{i,j,\cdots\})=\int[D\f]_M \exp[iS(\f_i,g_a)]\f_i(x_1) \f_j(x_2)\cdots, \label{eq:Zintegral}
\eeq
where $[D\f]$ is the path integral measure, and the subscript $M$ in $[D\f]_M$ means that the higher momemtum modes in the path integral 
are cut off at the scale $M$. $S$ is the action of the theory.

Let us change the parameters according to their mass dimensions as $M \to M'=e^\a M$ and $g_a \to g'_a =e^{ d_a \a}g_a $, with $\a$ an
infinitesimal real parameter. Then, we obtain
\beq
Z(g'_a, M', \{i,j,\cdots\}) &=& \int[D\f]_{M'} \exp[iS(\f_i,g'_a)]\f_i(x_1) \f_j(x_2)\cdots, \nonumber \\
&=&\int[D\f']_{M'} \exp[iS(\f'_i,g'_a)]\f'_i(x_1) \f'_j(x_2)\cdots,
\eeq
where in the second line we have changed the name of the integration variables from $\f$ to $\f'$. This is a trivial renaming and does not affect anything.
Now, we change the integration variables as
\beq
\f'_i(x)=e^{\D^c_i \a}\f(e^\a x).
\eeq
This is a nontrivial change of the variables.

Because the classical action is dimensionless, it is invariant under the change $g_a \to g'_a$ and $\f \to \f'$, that is,
\beq
S(g'_a,\f'_i)=S(g_a,\f_i).
\eeq
$\f_i(x_1),\f_j(x_2),\cdots$ appearing in Eq.~(\ref{eq:Zintegral}) is changed as, e.g.
\beq
\f'_i(x_1)=\f(x_1)+\a \left(x_1^\m \q_\m \f_i(x_1)+ \D^c_i\f_i(x_1)\right)+{\cal O}(\a^2),
\eeq
where we have expanded in terms of $\a$.

The most important point is that the path integral measure is also invariant in the sense that
\beq
\int [D\f']_{M'}=\int [D\f]_M, \label{eq:invmeasure}
\eeq
that is, it is invariant under the simaltaneous change of $\f$ and $M$~\footnote{If we change the variables $\f_i$ with $M$ fixed, we would get an anomalous 
Jacobian, which is studied in detail in Ref.~\cite{ArkaniHamed:1997mj}.}. 
What we are imagining here is an explicit cutoff of higher modes which may be
schematically represented as (e.g. for the case of scalar bosons with mass dimensions one)
\beq
\int [D\f]_M \sim \prod_{n_0,n_1,n_2,n_3:{\rm integer}}\frac{1}{2M}\int_{-M}^M d \f \left(x^\mu=\{an_0,an_1,an_2,an_3\} \right) ~~~(a\sim M^{-1}).
\eeq
With such a definition of the path integral, Eq.~(\ref{eq:invmeasure}) can be explicitly checked.
Of course, we do not know any rigorously defined Wilsonian path integral which maintains gauge invariance, supersymmetry, and other properties
which should be satisfied in quantum field theory.  
In this paper we simply assume the existence of a regularization scheme which satisfies Eq.~(\ref{eq:invmeasure}).
Note that the above discussion would become more complicated if we renormalize $\f_i$ as we change 
the renormalization scale $M$.

Using the invariance of the path integral measure and the action, we obtain
\beq
Z(g'_a, M', \{i,j,\cdots\}) &=& Z(g_a,M,\{i,j,\cdots\}) \nonumber \\
&&+\int[D\f]_M \exp[iS(\f_i,g_a)]\a \left(x_1^\m \q_\m \f_i(x_1)+ \D^c_i\f_i(x_1)\right) \f_j(x_2)\cdots \nonumber \\ &&+\cdots. \label{eq:dilat}
\eeq
On the other hand, from the renormalization group invariance, we obtain
\beq
Z(g'_a, M', \{i,j,\cdots\})=Z(g'_a-\a \b_a +{\cal O}(\a^2),M,\{i,j,\cdots\})
\eeq
where $\b_a$ is the $\b$ function of $g_a$,
\beq
\b_a=M\frac{\q}{\q M} g_a.
\eeq
Note that we have again used the invariance of the normalization of $\f_i$ with the change of $M$.
Thus, we obtain
\beq
Z(g'_a, M', \{i,j,\cdots\})&=& Z(g_a,M,\{i,j,\cdots\}) \nonumber \\
&&+\a \sum_a (d_ag_a-\b_a)\frac{\q}{\q g_a}Z(g_a,M,\{i,j,\cdots\})+{\cal O}(\a^2), \label{eq:rgchange1}
\eeq
where the derivative of $Z(g_a,M,\{i,j,\cdots\})$ can be calculated as
\beq
\frac{\q}{\q g_a}Z(g_a,M,\{i,j,\cdots\})=\int[D\f]_M \exp[iS(\f_i,g_a)] i \frac{\q S(\f_i,g_a)}{\q g_a}\f_i(x_1) \f_j(x_2)\cdots. \label{eq:rgchange2}
\eeq
Using Eqs.~(\ref{eq:dilat},\ref{eq:rgchange1},\ref{eq:rgchange2}), we finally obtain the formula
\beq
0&=&\vev{ \int d^4 yA_{\rm D}(y) \f_i(x_1) \f_j(x_2)\cdots} \nonumber \\
&&+ \vev{i \left(x_1^\m \q_\m \f_i(x_1)+ \D^c_i\f_i(x_1)\right) \f_j(x_2)\cdots} \nonumber \\ &&+\cdots, \label{eq:intdilatationward}
\eeq
where we have defined 
\beq
A_{\rm D}= \sum_a(d_ag_a-\b_a)\frac{\q {\cal L}(\f_i,g_a)}{\q g_a}. \label{eq:dilanom}
\eeq
${\cal L}$ is the Lagrangian of the theory.

The expression $i (x_1^\m \q_\m \f_i+ \D^c_i\f_i)$ is an infinitesimal version of the (classical) dilatation transformation.
Thus, we can expect that Eq.~(\ref{eq:intdilatationward}) is the integration of the following anomalous Ward identities,
\beq
\vev{\q_\m j^\m_{\rm D}(y) \f_i(x_1) \f_j(x_2)\cdots}&=&\vev{A_{\rm D}(y) \f_i(x_1) \f_j(x_2)\cdots} \nonumber \\
&&+ \d^4(y-x_1)\vev{i \left(x_1^\m \q_\m \f_i(x_1)+ \D^c_i\f_i(x_1)\right) \f_j(x_2)\cdots} \nonumber \\ &&+\cdots+({\rm total~derivatives}), \label{eq:localdilatation}
\eeq
where $j_{\rm D}^\mu$ is the current of dilatation, which is given in terms of the energy-momentum tensor as
\beq
j_{\rm D}^\m=x_\n T^{\n\m}.
\eeq
Because $\q_\m j_{\rm D}^\m=T^\mu_\mu$ up to contact terms, we can see that $A_{\rm D}$ given in Eq.~(\ref{eq:dilanom}) is the trace anomaly
up to total derivative, $T^\mu_\mu=A_{\rm D}+({\rm total~derivative})$.

Now we apply the above formulae to SUSY theories. The Wilson effective Lagrangian of the SUSY theory discussed in Section~\ref{sec:2} is 
given by
\beq
{\cal L}=\int d^4\h \sum_r Z_r \F_r^\dagger e^{-2V}\F_r+\int d^2\h\left( W(\F) + \frac{1}{4g_h^2}W^\a W_\a \right)+{\rm h.c.},
\eeq
where $Z_r$ is the wave function renormalization factor of $\F_r$, and $g_h$ is the holomorphic gauge coupling constant which runs only at the one-loop level
in renormalization group. We have neglected higher dimensional operators which are suppressed by the cutoff scale $M$.
We define the anomalous dimension of the fields as
\beq
\g_r=-M \frac{\q}{\q M}\log Z_r.
\eeq
Using the nonrenormalization theorem of the superpotential and the one-loop exactness of the holomorphic coupling constant,
Eq.~(\ref{eq:dilanom}) now reads 
\beq
A_{\rm D} &=& \int d^4\h \sum_r \g_r Z_r \F_r^\dagger e^{-2V}\F_r \nonumber \\
&&+\int d^2\h\left( 3W(\F)-\sum_r \F_r \frac{\q W}{\q \F_r} - \frac{3t(A)-\sum_r t(r)}{32\pi^2}W^\a W_\a \right)+{\rm h.c.} \nonumber \\
&=& -\frac{1}{2}\int d^4\h J  +\frac{3}{4} \int d^2\h X +{\rm h.c.},
\eeq
where $J$ and $X$ are defined by Eqs.~(\ref{eq:anompart},\ref{eq:correctionpart}). (Note that the fields are canonically normalized in Section~\ref{sec:2}
so that $Z_r=1$.)

After doing the $\h$ integral, the trace anomaly we finally obtain is (see \ref{app:A} for notation),
\beq
T^\m_\m &=& A_{\rm D}+({\rm total~derivative}) \nonumber \\ &=&-\frac{1}{4}D_J+\frac{3}{2} {\rm Re}F_X+({\rm total~derivative}),
\eeq
which agrees with Eq.~(\ref{eq:traceanom}) up to total derivative. 

We have some comments on the above results. First, we should expect that the total derivative term can not be determined by the above 
general argument, since the improvement Eq.~(\ref{eq:improvement}) is allowed in the definition of the energy-momentum tensor.
The next comment is on the classical scaling dimensions $\D^c_r$ appearing in the anomalous Ward identity Eq.~(\ref{eq:localdilatation}).
This scaling dimension is modified by the use of the Konishi anomaly equation~(\ref{eq:konishiequation}), because of the existence of contact terms
in that equation. Defining $Y$ as the right hand side of the Konishi anomaly equation
\beq
\frac{1}{4}\bar{D}^2J=Y \equiv \sum_r \left( -2\g_r \F_r\frac{\q W}{\q \F_r}-\frac{\g_r t(r)}{8\pi^2}W^\a W_\a \right),
\eeq
we have a supersymmetric extension of anomalous Ward identities (which can be derived as in the case of the usual anomalous Ward identities),
\beq
\vev{ (D_J(y)+2{\rm Re}F_Y(y)) \F_r(x_1,\h,\hh) \cdots }&=&-2i\d^4(y-x_1)\g_r \vev{\F_r(x_1,\h,\hh)\cdots} \nonumber \\ &&+\cdots+({\rm total~derivative}).
\eeq
After using the Konishi anomaly equation, the above contact terms modify the scaling dimensions as
\beq
\D^c_r \to \D_r=\D^c_r+\frac{\g_r}{2},
\eeq
which is consistent with the discussion of Subsection~\ref{sec:3.1}.

The last comment is on the role played by the holomorphic gauge coupling constant. 
The anomaly puzzle stated in the Introduction comes from higher order corrections to the $\b$ function,
and hence there is no puzzle if the $\b$ function is one-loop exact. The holomorphic gauge coupling indeed has the one-loop exact $\b$ function.
The above derivation of the trace anomaly utilizes the one-loop exactness of the holomorphic gauge coupling, confirming  
the idea in Ref.~\cite{Shifman:1986zi}.

\section*{Acknowledgements}
The author would like to thank K.-I.~Izawa, T.~Kugo, Y.~Nakayama and T.~T.~Yanagida for useful discussions.
This work was supported by 
World Premier International Research Center Initiative
(WPI Initiative), MEXT, Japan,
and supported
in part by JSPS Research Fellowships for Young Scientists.

\appendix
\setcounter{equation}{0}
\renewcommand{\theequation}{\Alph{section}.\arabic{equation}}
\renewcommand{\thesection}{Appendix~\Alph{section}}

\section{Notation and Convention}\label{app:A}
\setcounter{equation}{0}
For superspace manipulation, we follow the conventions of Wess and Bagger~\cite{Wess:1992cp},
except that a matter kinetic term is given by $\F^\dagger e^{-2V}\F$, and a gauge field strength chiral field is given by
$W_\a=\frac{1}{8}\bar{D}^2 e^{2V}D_\a e^{-2V}$ where $V$ is a vector superfield for a gauge multiplet.
As in Ref.~\cite{Komargodski:2010rb}, a vector $\ell_\m$ is often expressed in bi-spinor notation as $\ell_{\a\aa}=-2\s^\mu_{\a\aa}\ell_\mu$, 
$\ell_\mu=\frac{1}{4}\ss_\mu^{\aa\a}\ell_{\a\aa}$. 

A chiral field $X$ is given as
\beq
X=\f_X(y)+\sqrt{2}\h\p_X(y)+\h^2F_X(y),~~~y=x+i\h\s^\mu \hh.
\eeq
A real vector superfield $J$ is given as
\beq
J&=&C_J+i\h\chi_J-i\hh\bar{\chi}_J+\frac{i}{2}\h^2L_J-\frac{i}{2}\hh^2\bar{L}_J+\h\s_\mu\hh j_J^\mu \nonumber \\
&&+i\h^2\hh\left[\bar{\l}_J+\frac{i}{2}\ss^\m \chi_J \right]-i\hh^2\h\left[\l_J+\frac{i}{2}\s^\m \bar{\chi}_J \right]+\frac{1}{2}\h^2\hh^2\left[D_J+\frac{1}{2}\q^2C_J\right].
\eeq
Using these notations, the solution of Eq.~(\ref{eq:supercurrenteq}) is represented as~\cite{Komargodski:2010rb}
\beq
\J^\mu &=& j_R^\mu+\h\left(S^\mu-\frac{1}{\sqrt{2}}\s^\mu\bar{\p}_X\right)+\hh\left(\bar{S}^\mu+\frac{1}{\sqrt{2}}\ss^\mu \p_X \right)
+\frac{i}{2}\h^2\q^\mu \bar{\f}_X-\frac{i}{2}\hh^2\q^\mu\f_X \nonumber \\
&&+\h\s_\nu\hh\left(2T^{\m\n}-\y^{\m\n}{\rm Re}F_X+\frac{1}{4}\e^{\m\n\r\s}\left(2\q_\r j_{R\s} -\q_\r j_{J\s}\right) \right) \nonumber \\
&&+\h^2\left( \frac{i}{2}\q_\r S^\mu \s^\r-\frac{i}{2\sqrt{2}}\q_{\r} \bar{\p}_X\ss^\r\s^\m \right)\hh+\hh^2 \h\left(-\frac{i}{2}\s^\r \q_\r \bar{S}^\mu  
+\frac{i}{2\sqrt{2}}\s^\m \ss^\r \q_{\r} \p_X\right) \nonumber \\
&&+\h^2\hh^2\left(\frac{1}{2}\q^\m\q_\n j_R^\n -\frac{1}{4} \q^2j_R^\m     \right),
\eeq
where $j_R^\mu$ is an R-symmetry current, $S_\a^\mu$ a supersymmetry current, and $T^{\m\n}$ an energy-momentum tensor.

In a redefinition of a supercurrent, we encounter an expression of the form $[D_\a,\bar{D}_\aa]U$ with $U$ a real vector superfield. In components, it is given as
\beq
\frac{1}{4}\ss^{\mu\aa\a}[D_\a,\bar{D}_\aa]U&=&-j_U^\mu+\h(i\s^m\bar{\l}_U+\q^m\chi_U)+\hh(i\ss^m \l_U+\q^\m\bar{\chi}_U)
+\frac{1}{2}\h^2\q^\mu L_U+\frac{1}{2}\hh^2\q^\mu \bar{L}_U \nonumber \\
&&+\h\s_\n\hh\left(D_U \y^{\m\n}+\q^\m\q^\n C_U+\e^{\m\n\r\s}\q_{\r}j_{U\s}\right) \nonumber \\
&&+\frac{1}{2}\hh^2\h\left(-\s^\m\ss^\n\q_\n\l_U+i\s_\n\q^\m\q^\n\bar{\chi}_U\right)+\frac{1}{2}\h^2\hh\left(-\ss^\m\s^\n\q_\n\bar{\l}_U
+i\ss_\n\q^\m\q^\n\chi_U\right) \nonumber \\
&&+\h^2\hh^2\left(-\frac{1}{2}\q^\m\q_\n j_U^\n+\frac{1}{4}\q^2 j_U^\m \right).
\eeq

\section{Non-conserved currents and anomalous Ward identities} \label{app:B}
\setcounter{equation}{0}
A conserved current can be characterized by Ward identities it satisfies. Suppose that a theory has fields $\f_i$ with charge $q_i$ under some symmetry,
and the current of the symmetry is denoted by $j^\mu$.
Then, correlation functions satisfy the following Ward identities~\footnote{We neglect Schwinger terms for simplicity.}:
\beq
\vev{\q_\m j^\mu(y) \f_{i}(x_1) \f_{j}(x_2)\cdots \f_k(x_3)}&=&q_i\d^4(y-x_1)\vev{\f_i(x_1)\f_j(x_2)\cdots\f_k(x_3)}\nonumber \\
&&+q_j\d^4(y-x_2)\vev{\f_i(x_1)\f_j(x_2)\cdots \f_k(x_3)} \nonumber \\ &&+\cdots \nonumber \\
&&+q_k\d^4(y-x_3)\vev{\f_i(x_1)\f_j(x_2)\cdots \f_k(x_3)},
\eeq
where $\vev{\cdots}$ means path integral expectation values. There are contact terms proportional to the delta function, and the existence of these contact terms
is the defining property of the current.

We should consider contact terms for the case of non-conserved currents as well. We define a non-conserved current as follows. 
We say that a non-conserved current $j^\m$ with charge assignment $q_i$ to fields $\f_i$
has anomaly $A$ (which is not necessarily a quantum anomaly, but also contains tree-level violation of the conservation of the current) if 
it satisfies anomalous Ward identities,
\beq
\vev{\q_\m j^\mu(y) \f_{i}(x_1) \f_{j}(x_2)\cdots \f_k(x_3)}&=&\vev{A(y)\f_i(x_1)\f_j(x_2)\cdots\f_k(x_3)} \nonumber \\
&&+q_i\d^4(y-x_1)\vev{\f_i(x_1)\f_j(x_2)\cdots\f_k(x_3)}\nonumber \\
&&+q_j\d^4(y-x_2)\vev{\f_i(x_1)\f_j(x_2)\cdots \f_k(x_3)} \nonumber \\ &&+\cdots \nonumber \\
&&+q_k\d^4(y-x_3)\vev{\f_i(x_1)\f_j(x_2)\cdots \f_k(x_3)}. \label{eq:anomalousward}
\eeq
The usual anomaly equation $\q_\m j^\m=A$ is valid only up to contact terms, and the contact terms are specified by the charge assignment.

Let us take another non-conserved current $j'^\mu$ and define $j_{\rm new}^\m=j^\m+j'^\m$. Then, $j_{\rm new}^\mu$ is a current
with charge assignment $q_i$ to fields $\f_i$ with anomaly $A+\q_\m j'^\m$, because it trivially satisfies
\beq
\vev{\q_\m j_{\rm new}^\mu(y) \f_{i}(x_1) \f_{j}(x_2)\cdots \f_k(x_3)}&=&\vev{(A(y)+\q_\m j'^\m(y))\f_i(x_1)\f_j(x_2)\cdots\f_k(x_3)} \nonumber \\
&&+q_i\d^4(y-x_1)\vev{\f_i(x_1)\f_j(x_2)\cdots\f_k(x_3)}\nonumber \\
&&+q_j\d^4(y-x_2)\vev{\f_i(x_1)\f_j(x_2)\cdots \f_k(x_3)} \nonumber \\ &&+\cdots \nonumber \\
&&+q_k\d^4(y-x_3)\vev{\f_i(x_1)\f_j(x_2)\cdots \f_k(x_3)}.
\eeq
Note that the charge assignment is not changed at this point. If $j'^\m$ is a current with charge assignment $q'_i$ to fields $\f_i$ and has anomaly $A'$,
we can also interpret $j_{\rm new}^\m$ as the current with charge assignment $q_i+q'_i$ to fields $\f_i$ and anomaly $A+A'$.
These two interpretations are completely consistent with each other.
The charge assignment of non-conserved currents has this ambiguity, and a non-conserved current becomes meaningful
only after specifying both the charge assignment and the anomaly of the current.
For example, the most trivial statement is that an arbitrary vector $\ell^\mu$ is a current with charge assignment $0$ to all fields
with anomaly $\q_\m \ell^\mu$. In all the situations considered in this paper, the above $j'^\m$ is suppressed by coupling constants compared with $j^\m$, so 
in the language of renormalization, the difference between $j^\m$ and $j^\m_{\rm new}$, $j'^\mu$, can be seen as a finite counterterm to the operator.

In the case of the usual Adler-Bell-Jackiw anomaly, $A$ is given by a topological term, $A \propto F^{\m\n}\tilde{F}_{\m\n}$.
In our opinion, the real significance of the Adler-Bardeen theorem is not that the relative coefficient of the current $j^\mu$ and 
the anomaly $F^{\m\n}\tilde{F}_{\m\n}$ is one-loop exact, but that the relative coefficient of the contact terms and $F^{\m\n}\tilde{F}_{\m\n}$ in the 
anomalous Ward identities Eq.~(\ref{eq:anomalousward}) is one-loop exact. If we integrate Eq.~(\ref{eq:anomalousward}) in terms of $y$, 
the left hand side is an integration of total derivative and hence vanishes (if there are no massless Nambu-Goldstone-like poles).
We obtain
\beq
0&=&\vev{ \left(\int d^4 yA(y) \right) \f_i(x_1)\f_j(x_2)\cdots\f_k(x_3)} \nonumber \\
&&+(q_i+q_j+\cdots+q_k) \vev{\f_i(x_1)\f_j(x_2)\cdots\f_k(x_3)}
\eeq
At least in the case of non-Abelian gauge theories, the integral of $A \propto F^{\m\n}\tilde{F}_{\m\n}$ is quantized in instanton calculus, and hence the relative coefficient of $A$ and contact terms should not receive higher order corrections to satisfy $\int d^4 y A(y)+\sum q=0$ at some instanton background 
(see e.g. Ref.~\cite{Coleman} for a lucid review). In this point of view, it does not matter whether $A$ is given by
$ F^{\m\n}\tilde{F}_{\m\n}$ or $ F^{\m\n}\tilde{F}_{\m\n}+\q_\m j'^\m$, because the integration of the total derivative vanishes.
For example, as pointed out in Section~\ref{sec:2}, in SUSY gauge theories the imaginally part of the $F$-term of $W^\a W_\a$ (which gives anomaly)
is not proportional to $F^{\m\n}\tilde{F}_{\m\n}$ but 
\beq
{\rm Im} \left( W^\a W_\a|_{\h^2} \right) = -\frac{1}{2}F^{\m\n}\tilde{F}_{\m\n}-\q_\mu (\l \s^\mu \bar{\l}),\label{eq:imgaugestrength}
\eeq
where $\l$ is the gaugino. The presence of $\q_\mu (\l \s^\mu \bar{\l})$ is one of the sources of the confusion about the anomaly puzzle in the past
(see e.g. Ref.~\cite{Jones:1983ip}).
But in the context of the present discussion, this term does not spoil the Adler-Bardeen theorem since it is a total derivative.

Perturbative calculations also suggest~\cite{Larin:1993tq} that the Adler-Bardeen theorem is correct in the above sense.
Contrary to the case of conserved currents, 
non-conserved currents usually receive quantum corrections and are renormalized, $j^\mu \to Z_j j^\mu$ where $Z_j$ is the renormalization constant.
The anomaly $A \sim F^{\m\n}\tilde{F}_{\m\n}$ is renormalized as~\cite{Breitenlohner:1983pi} $A \to A+Z_{Aj}\q_\mu j^\mu$.
If we appropriately normalize the anomalous current so that the Adler-Bardeen theorem in the usual sense (i.e. one-loop exactness of the relative
coefficient between $j^\m$ and $F^{\m\n}\tilde{F}_{\m\n}$)
is maintained in the renormalization group flow,
we have $Z_{Aj}=Z_j-1$, and Eq.~(\ref{eq:anomalousward}) is maintained. However, even if we do not normalize the current appropriately,
the relative coefficient of the contact terms and the topological term $F^{\m\n}\tilde{F}_{\m\n}$ is not changed under the renormalization group, 
and hence the Adler-Bardeen theorem in the above sense is still correct.   

In this paper, we usually define non-conserved currents ``appropriately'' (by using the ambiguity $j^\m \to j^\m+j'^\m$ and $A \to A+\q_\m j'^\m$ discussed above) 
so that the supersymmetric version of the Adler-Bardeen theorem in the usual sense
is maintained. However, the ``appropriate'' definition of currents depends on the number of SUSY, ${\cal N}$, as follows:
\beq
\q_\m j^\m \sim \left\{
\begin{array}{lc}
F^{\m\n}\tilde{F}_{\m\n}&~~({\cal N}=0),\\
F^{\m\n}\tilde{F}_{\m\n}+2\q_\mu (\l \s^\mu \bar{\l})&~~({\cal N}=1),\\
F^{\m\n}\tilde{F}_{\m\n}+2\q_\mu (\l \s^\mu \bar{\l}+\p_{Ad} \s^\mu \bar{\p}_{Ad}+i(D^\m \f_{Ad}^\dagger)\f_{Ad}-i\f_{Ad}^\dagger(D^\mu \f_{Ad})) &~~({\cal N}=2),
\end{array} 
\right. \nonumber
\eeq
where $\f_{Ad}$ and $\p_{Ad}$ are the scalar and fermion components of the adjoint chiral field $\F_{Ad}$ in ${\cal N}=2$ theory
(see Eq.~(\ref{eq:imgaugestrength}) and Subsection~\ref{sec:3.3} ). These definitions of the anomaly only differ
in the total derivative terms, and hence the difference is not significant for the Adler-Bardeen theorem in the sense discussed above. 

There is in fact a nontrivial prediction of the above consideration combined with the discussion of Section~\ref{sec:2}. 
Let us consider ${\cal N}=1$ pure Yang-Mills.
The anomaly equation of the R-symmetry current is given by
\beq
\q_\mu j_R^\mu=\frac{t(A)}{8\pi^2}\left(\frac{1}{2}F^{\m\n}\tilde{F}_{\m\n}+\q_\mu (\l \s^\mu \bar{\l}) \right).
\eeq
We do not kwon (without further detailed study) the precise normalization of the operator $\l \s^\m \bar{\l}$, because of the ambiguity of the finite renormalization in perturbative calculations. 
However, at least at the leading order, it is related to the R-symmetry current as
\beq
\l \s^\mu \bar{\l}=-(g^2+{\cal O}(g^4))j_R^\mu.
\eeq
(Note that the normalization of the gauge kinetic term is given by $\int d^2 \h (1/4g^2)W^\a W_\a$.)
Thus, we can rewrite the anomaly equation as
\beq
\q_\m j_{{\cal N}=0}^\m &=& \frac{t(A)}{16\pi^2}F^{\m\n}\tilde{F}_{\m\n}, \\
 j_{{\cal N}=0}^\m &\equiv&\left(1+\frac{t(A)}{8\pi^2}g^2+{\cal O}(g^4)\right) j_R^\mu.\label{eq:N0current}
\eeq
The current $ j_{{\cal N}=0}^\m$ is the one which is normalized so as to satisfy the Adler-Bardeen theorem in the ${\cal N}=0$ description.
Now, the important point is that the current $j_R^\mu$ is not multiplicatively renormalized, because it is in the same multiplet with the energy-momentum
tensor, and it does not mix with other currents simply because there are no other currents in pure Yang-Mills theory. 
Thus, $j_R^\m$ is renormalization group invariant. Then, from Eq.~(\ref{eq:N0current}), we can see that the anomalous dimension
of the current $ j_{{\cal N}=0}^\m$ is predicted as
\beq
\g_{j,{\cal N}=0} &=& \m \frac{\q}{\q \m}\log \left(1+\frac{t(A)}{8\pi^2}g^2+{\cal O}(g^4)\right) \nonumber \\
&=&\left(\frac{g^2}{16\pi^2}\right)^2 \left(-12 t(A)^2\right)+{\cal O}(g^6).
\eeq
This result indeed agrees~\footnote{To compare the result with Ref.~\cite{Kodaira:1979pa,Larin:1993tq},
one should note the difference of the definition of the anomalous dimension and should also note that
we are considering a Majorana adjoint fermion (i.e. gaugino) instead of Dirac fermions.} with the explicit two-loop calculation~\cite{Kodaira:1979pa,Larin:1993tq}.
We leave more detailed study on these points for future work.

\end{document}